\documentclass{ptephy_v1}

\preprintnumber{KEK-TH-2784} 

\usepackage{graphicx}

\usepackage{amsmath,amssymb,bm,amsfonts}

\usepackage[utf8]{inputenc}
\usepackage{verbatim}

\numberwithin{equation}{section}

\usepackage{cancel}

\newcommand{\be}{\begin{equation}}
\newcommand{\ee}{\end{equation}}
\newcommand{\bea}{\begin{eqnarray}}
\newcommand{\eea}{\end{eqnarray}}

\newcommand{\eq}[1]{Eq.~(\ref{eq:#1})}
\newcommand{\sect}[1]{Sec.~\ref{sec:#1}}
\newcommand{\appen}[1]{Appendix~\ref{sec:#1}}

\newcommand{\del}{\partial}

\newcommand{\bra}{\langle}
\newcommand{\ket}{\rangle}
\newcommand{\calO}{{\cal O}}

\newcommand{\example}{{\it e.g.}}

\newcommand{\bh}{black hole\ }

\bmdefine{\bmq}{{\bm{q}}}
\bmdefine{\bmk}{{\bm{k}}}
\bmdefine{\bmx}{{\bm{x}}}
\bmdefine{\bmy}{{\bm{y}}}
\bmdefine{\bmr}{{\bm{r}}}
\bmdefine{\bmnabla}{{\bm{\nabla}}}
\bmdefine{\bmA}{ \bm{A} }
\bmdefine{\bmD}{ \bm{D} }

\bmdefine{\bmPhi}{ \bm{\Phi} }
\bmdefine{\bmPsi}{ \bm{\Psi} }
\bmdefine{\bmcalO}{ \bm{\mathcal{O}} }
\bmdefine{\bmrho}{ \bm{\rho} }
\bmdefine{\bmX}{{\bm{X}}}

\newcommand{\vecx}{\vec{x}}
\newcommand{\vecy}{\vec{y}}

\newcommand{\epsmu}{\epsilon_{\mu}}

\newcommand{\nq}{\mathfrak{q}}

\newcommand{\nw}{\mathfrak{w}}

\bmdefine{\bmg}{{\bm{g}}}
\bmdefine{\bmR}{{\bm{R}}}

\newcommand{\mfh}{\mathfrak{h}}

\newcommand{\half}{\frac{1}{2}}

\newcommand{\vecX}{\vec{X}}

\newcommand{\mfa}{\mathfrak{a}}

\newcommand{\Cone}{\delta\psi}

\newcommand{\SC}{superfluid}
\newcommand{\SCs}{superfluids}
\newcommand{\HSC}{holographic superfluid}
\newcommand{\HSCs}{holographic superfluids}

\newcommand{\thetarho}{(\bm\rho \Theta)}
\newcommand{\Zin}{Z_\text{in}}

\newcommand{\eigen}{\alpha}

\newcommand{\tin}{\text{in}}
\newcommand{\tout}{\text{out}}
\newcommand{\tI}{\text{I}}
\newcommand{\tO}{\text{O}}
\newcommand{\up}{\vec{u}}
\newcommand{\down}{\vec{d}}

\begin{document}

\title{Pole-skipping without master variable and holographic superfluids}%

\author[1]{Makoto Natsuume}
\affil[1]{KEK Theory Center, Institute of Particle and Nuclear Studies, High Energy Accelerator Research Organization, Tsukuba, Ibaraki, 305-0801, Japan 
\thanks{Also at 
Department of Particle and Nuclear Physics, 
SOKENDAI (The Graduate University for Advanced Studies), 1-1 Oho, 
Tsukuba, Ibaraki, 305-0801, Japan;
Department of Mechanical Engineering, Mie University, 
Tsu, 514-8507, Japan.} 
\email{makoto.natsuume@kek.jp}}

\author[2]{Takashi Okamura}
\affil[2]{Department of Physics and Astronomy, Kwansei Gakuin University, Sanda, Hyogo, 669-1337, Japan \email{tokamura@kwansei.ac.jp}}

\begin{abstract}%
The pole-skipping is a universal property of Green's functions at strong coupling found by the AdS/CFT duality. There is a conventional formalism of the pole-skipping, 
but it relies on the existence of a ``master variable." Namely, it is applicable to a system with a single field. We propose an alternative formalism that does not rely on a master variable. As an example, we study the pole-skipping of holographic superfluids. A ``hydrodynamic" pole such as the diffusion pole is usually regarded as a pole-skipping point. But we point out that not all hydrodynamic poles are pole-skipping points 
because their residues may not vanish.
\end{abstract}

\subjectindex{AdS/CFT correspondence, Black holes in string theory}



\maketitle


\section{Introduction}

In the AdS/CFT duality or holographic duality \cite{Maldacena:1997re,Witten:1998qj,Witten:1998zw,Gubser:1998bc}, one often encounters universal relations in the strong coupling limit.%
\footnote{
See, \example, Refs.~\cite{CasalderreySolana:2011us,Natsuume:2014sfa,Ammon:2015wua,Zaanen:2015oix,Hartnoll:2016apf,Baggioli:2019rrs} for AdS/CFT textbooks.}
For example,
\begin{enumerate}
\item
The most famous one is $\eta/s=1/(4\pi)$ where $\eta$ is the shear viscosity and $s$ is the entropy density \cite{Kovtun:2004de}. 
\item
Another example is many-body quantum chaos: The Lyapunov exponent is $\lambda_L=2\pi T$ \cite{Shenker:2013pqa,Roberts:2014isa,Roberts:2014ifa,Shenker:2014cwa,Maldacena:2015waa}. 
\item
The pole-skipping phenomenon is one another example \cite{Grozdanov:2017ajz,Blake:2018leo,Grozdanov:2019uhi,Blake:2019otz,Natsuume:2019xcy}%
\footnote{See, \example, Refs.~\cite{Natsuume:2019sfp,Natsuume:2019vcv,Wu:2019esr,Balm:2019dxk,Ceplak:2019ymw,Liu:2020yaf,Ahn:2019rnq,Ahn:2020bks,Abbasi:2020ykq,Jansen:2020hfd,Ramirez:2020qer,Ahn:2020baf,Natsuume:2020snz,Kim:2020url,Sil:2020jhr,Abbasi:2020xli,Ceplak:2021efc,Jeong:2021zhz,Natsuume:2021fhn,Blake:2021hjj,Kim:2021xdz,Wang:2022mcq,Amano:2022mlu,Yuan:2023tft,Grozdanov:2023txs,
Natsuume:2023lzy,Natsuume:2023hsz,Abbasi:2023myj,Ning:2023ggs,Cartwright:2024iwc,Grozdanov:2025ner,Lu:2025jgk,Ahn:2025exp,Grozdanov:2025ulc,Davison:2025xdj}
for some other works of the pole-skipping.}
 and it partly includes the universal Lyapunov exponent. 
\end{enumerate}
These ``universalities" all come from the universal nature of the \bh horizon physics. 

Consider the field perturbation of the form $e^{-i\omega t+iqx}$, \ie, $\omega$ is frequency and $q$ is wave number. 
In the pole-skipping phenomenon, finite-temperature Green's functions are not uniquely determined at special points  in the complex momentum space $(\omega,q)$.
Green's functions $G^R$ have the structure
\begin{align}
G^R=\frac{0}{0}~.
%
\end{align}
Namely, the residue of a pole vanishes. 

This phenomenon itself occurs even in elementary quantum mechanics \cite{Natsuume:2021fhn}. But the important point is that there is a  universality for the pole-skipping points $\omega$. The pole-skipping points start from
\begin{align}
\nw :=\frac{\omega}{2\pi T}=(s-1)i~,
%
\end{align}
and continue to $\nw_n = (s-1-n)i $ for a non-negative integer $n$ ($T$ is temperature and $s$ is the spin of the bulk field). 
For example, 
\begin{itemize}
\item
For a scalar field, they start from $\nw_1=-i$. 
\item
For the Maxwell field, they start from $\nw_0=0$ which is a hydrodynamic pole. 
\item
For the gravitational sound mode, they start from $\nw_{-1}=+i$. It is argued that the $\nw_{-1} = +i$ point is related to many-body quantum chaos.
\end{itemize}

There is a conventional formalism of pole-skipping proposed in Ref.~\cite{Blake:2019otz}. However, this formalism is applicable to a system with a single field. In general, one would like to study a system with multiple number of fields. In some cases, one can find a ``master variable," and the system reduces to a single field equation. Then, one can use the conventional formalism. 

However, it is in general very difficult to find a master variable. 
Its existence is not even guaranteed. 
One often spends most of time to find a master variable in a pole-skipping analysis. 
Here, we propose an alternative formalism that does not rely on the master variable. The formalism was partly developed in our early work \cite{Natsuume:2023lzy}. In this paper, we develop the formalism further and apply it to \HSCs\ \cite{Gubser:2008px,Hartnoll:2008vx,Hartnoll:2008kx}. 

Holographic \SCs\ describe \SCs. Typically, a \HSC\ is an Einstein-Maxwell-complex scalar system. Even if one uses the ``probe limit," where the background geometry is fixed, one has to study a Maxwell-complex scalar system
 in an AdS \bh background. Such a system is hard to solve, and one usually needs a numerical computation. An analytic solution is available only for a particular bulk scalar mass in the hydrodynamic limit  \cite{Herzog:2010vz}.
 But we apply our formalism to a \HSC\ with \textit{arbitrary} bulk scalar mass.
 Using our formalism, we find pole-skipping points analytically.

It is interesting to study the pole-skipping for \HSCs.  A ``hydrodynamic pole" is usually regarded as a pole-skipping point. 
One example is the ``diffusion pole." 
The charge density Green's function typically takes the form
\begin{align}
G^R \propto \frac{q^2}{i\omega-Dq^2}~,
%
\end{align}
where $D$ is the diffusion constant. The Green's function has the $0/0$ structure at $(\omega,q)=(0,0)$.
Another examples are the ``shear pole" and the ``sound pole" in gravitational perturbations.  

Hydrodynamic poles represent macroscopic variables. One typical macroscopic variable is a conserved quantity. They are guaranteed to survive in the low-energy $\omega\to0$, long-wavelength limit $q\to0.$ For the sound, shear, and diffusion poles, the associated conserved quantities are the energy density, the momentum density, and the charge density. 

However, there are the other macroscopic variables. For a \SC,  a second-order phase transition occurs at the critical point, 
and the order parameter $\psi$ becomes massless there so that $\psi$ has a hydrodynamic pole. 
Then, we would like to address the question:
\begin{center}
Is a hydrodynamic pole always a pole-skipping point?
\end{center}
This question has never been answered in the literature. If the pole-skipping point of a scalar field always starts from $\nw_1=-i$, a hydrodynamic pole associated with a scalar field is not a pole-skipping point. However, there is no generic proof, in particular when a scalar field has a hydrodynamic pole as far as we are aware: it is difficult to apply the conventional formalism to a complicated system such as the \HSC. We show that there is no new pole-skipping point associated with the massless order parameter. 


One can easily show this if one approaches the critical point from the high-temperature phase (\sect{high_T}). This is because the complex scalar perturbation $\delta\Psi$ decouples from Maxwell perturbations. 

However, it is nontrivial if one approaches the critical point from the low-temperature phase (\sect{low_T}). The complex scalar perturbation couples with Maxwell perturbations, but a master variable is not known for \HSCs\ in general.
We use our formalism to show that there is no hydrodynamic pole-skipping associated with the massless order parameter.   
Then, a hydrodynamic pole can be pole-skipping point only for the pole associated with a conserved charge.
 
 We explain our formalism pedagogically with simple examples, the scalar field and the Maxwell field in \sect{formalism}.  Then, we apply it to \HSCs\ in \sect{HSC_pole_skip}. In \sect{formal}, we justify our formalism by a formal argument. See \sect{summary} for the summary of our formalism. We hope that our formalism will be helpful for future analysis of the pole-skipping.

 Incidentally, we consider holographic superfluids, but many results equally apply to holographic superconductors. 
 The difference between two systems lies in the difference of the boundary conditions for the Maxwell field at the asymptotic infinity. However, the pole-skipping analysis itself is based on the near-horizon physics (\sect{discussion}). 

\section{Matrix formalism and examples}\label{sec:formalism}

In this paper, we consider the pole-skipping in the Schwarzschild-AdS$_5$ (SAdS$_5$) black hole background:%
\footnote{We use upper-case Latin indices $M, N, \ldots$ for the 5-dimensional bulk spacetime coordinates and use Greek indices $\mu, \nu, \ldots$ for the 4-dimensional boundary coordinates. The boundary coordinates are written as  $x^\mu = (t, x^i) =(t, \vecx)=(t,x,y,z)$. }
\begin{subequations}
\label{eq:sads5}
\begin{align}
ds_5^2 &= \left( \frac{r}{L} \right)^2(-fdt^2+dx^2+dy^2+dz^2)+L^2\frac{dr^2}{r^2f} \\
&= \left( \frac{r_0}{L} \right)^2\frac{1}{u} (-fdt^2+dx^2+dy^2+dz^2)+L^2\frac{du^2}{4u^2f}~, \\
f &= 1-\left(\frac{r_0}{r}\right)^4 = 1-u^2~, 
%
\end{align}
\end{subequations}
where $u:=r_0^2/r^2$.  The Hawking temperature is given by $\pi T =r_0/L^2$. 
One can carry out a similar analysis for the SAdS$_{p+2}$ black hole, but we focus on the $p=3$ case for simplicity.
We also set the AdS radius $L=1$ and the horizon radius $r_0=1$, so we work in the unit $\pi T=1$.
In this paper, we consider the linear perturbations of the form 
\begin{align}
\varphi=\varphi(u)e^{-i\omega t+iqx}~.
%
\end{align}

As an example, consider the minimally-coupled scalar field:
\begin{subequations}
\label{eq:scalar}
\begin{align}
0&=(\nabla^2-m^2)\phi \\
&\propto u \left( \frac{f}{u}\phi' \right)' +\left[ \frac{\nw^2}{uf} -\frac{\nq^2}{u} - \frac{m^2}{4u^2} \right] \phi~,
%
\end{align}
\end{subequations}
where $\nw:=\omega/(2\pi T), \nq:=q/(2\pi T)$, and $':=\del_u$.

In AdS/CFT, one solves such a bulk field equation and obtains the dual Green's function. Because this is a second-order differential equation, one generally has 2 solutions. They are called the ``incoming-wave" solution and the ``outgoing-wave" solution at the horizon. One chooses the incoming wave to obtain the retarded Green's function. 
Similarly, one has 2 independent solutions at the asymptotic infinity $u=0$, but the incoming-wave boundary condition relates 2 solutions. 
Suppose that one obtains the bulk solution. 
For simplicity, we consider $m^2=-4$.
In this case, the asymptotic behavior at $u\to 0$ becomes
\begin{align}
\phi &\sim 
\frac{J}{2} u\ln u - \bra\calO\ket\, u+ \cdots~, \quad (u\to0)~.
%
\end{align}
According to the standard AdS/CFT dictionary, $\calO$ is the scalar operator that is dual to the bulk scalar field $\phi$ and $J$ is its source. From the linear response theory, the retarded Green's function is then given by
\begin{align}
G^R_{\calO\calO} &=-\frac{\del \bra\calO\ket}{\del J} ~.
\label{eq:Green}
\end{align}

The field equation \eqref{eq:scalar} has a regular singular point at the horizon $u=1$. 
In principle, one can construct the full solution by a power-series expansion:%
\footnote{
The power-series expansion is guaranteed to converge only inside a radius $\rho$ around $u=1$ in the complex $u$-plane, where $\rho$ is the distance to the nearest singular point. Because we construct the solution from $u=1$ to $u=0$, one needs $\rho \geq1$. The field equation has regular singular points at $u=0,\pm 1,\infty$, so the power-series expansion is guaranteed to converge at $u=0$.}
\begin{align}
\phi = \sum_{n=0}\phi_n(u-1)^{n+\lambda}~.
%
\end{align}
In a pole-skipping phenomenon, the bulk solution is not uniquely determined at pole-skipping points. As a result, the dual Green's function is not uniquely determined there.%
\footnote{Because we solve a homogeneous linear differential equation, there is always an ambiguity of an overall factor of the bulk solution. But this is not relevant to Green's functions because one takes the ratio of the fast falloff to the slow falloff \eqref{eq:Green}.}

In general, it is difficult to obtain the closed form of the solution. But it was pointed out that the nonuniqueness comes from the near-horizon behavior of the solution \cite{Grozdanov:2017ajz,Blake:2018leo,Grozdanov:2019uhi,Blake:2019otz,Natsuume:2019xcy}, so it is not necessary to obtain the complete solution. It is enough to study the power-series solution up to $n$ to find the pole-skipping points at $\nw=-in$. 

\subsection{Scalar field}

Now, let us explain our pole-skipping formalism using the above scalar field example. In a pole-skipping analysis, it is conventional to use the Eddington-Finkelstein coordinate system, but it is not necessary. We use the original coordinate system \eqref{eq:sads5}. Equivalently, one may impose the incoming-wave boundary condition ansatz
\begin{align}
\phi=f^{-i\nw/2}\Zin
%
\end{align}
and study the field equation for $\Zin$, but it is not necessary either. We impose the incoming-wave boundary condition later. In a general problem with many fields, the field equations with the ansatz become complicated, and it is easier to study original field equations (see \appen{method2} for the method using $\Zin$).

The field obeys a second-order differential equation. Rewrite them as 2 first-order differential equations. 
And rewrite them in a matrix form:
\begin{subequations}
\label{eq:matrix_scalar}
\begin{align}
0 &= \vecX'- M \vecX~, \\
\vecX &= \begin{pmatrix} 
\phi \\ 
f\phi'
 \end{pmatrix}~,
\\
M &= 
  \begin{pmatrix} 
  0 &\dfrac{1}{f} \\
  -\dfrac{\nw^2}{uf} +\dfrac{\nq^2}{u} - \dfrac{1}{u^2} & \dfrac{1}{u}
  \end{pmatrix}~.
%
\end{align}
\end{subequations}
Here, 
\begin{enumerate}
\item
We choose $\vecX$ so that both components have the same asymptotic behavior $(u-1)^{-i\nw/2}$ at the horizon $u\to1$.
\item
Then, the matrix $M$ diverges no more rapidly than $1/(u-1)$, and one can use the standard Frobenius method.
\end{enumerate}
The matrix $M$ is expanded as
\begin{subequations}
\begin{align}
M &= \frac{M_{-1}}{u-1}+M_0+M_1(u-1)+\cdots~,
\label{eq:M_exp} \\
M_{-1}&=\frac{1}{2}
  \begin{pmatrix} 
  0& -1 \\
  \nw^2 & 0
  \end{pmatrix}~,
\label{eq:M-1}
\end{align}
\end{subequations}
and the solution can be written as a power series:
\begin{align}
\vecX = \sum_{n=0}\, \vecx_n\, (u-1)^{n+\lambda}~.
\label{eq:X_series}
\end{align}
The key observation is that
\begin{center}
``The coefficient vector $\vecx_n$ becomes ambiguous at a pole-skipping point."
\end{center}
$\vecx_n$ are  the coefficients of the Frobenius series. 
When $\vecx_n$ is ambiguous, the bulk solution becomes ambiguous or is not uniquely determined.
As a result, the corresponding Green's function in the dual theory is not uniquely determined as well.
Substituting \eq{X_series} into the field equation \eqref{eq:matrix_scalar}, at the lowest order, one obtains
\begin{align}
0 &=(\lambda - M_{-1})\vecx_0~.
\label{eq:eigen}
\end{align}
This is the indicial equation for $\lambda$ and is the eigenvalue equation for $M_{-1}$. 
The eigenvalues and the eigenvectors of $M_{-1}$ are
%
%
\begin{subequations}
\begin{align}
\lambda&=-i\nw/2~, \quad 
\vecx_0= 
\begin{pmatrix}
 1 \\
 i\nw 
\end{pmatrix}
~,
\\
\lambda&=+i\nw/2~, \quad 
\vecx_0= 
\begin{pmatrix}
 1\\
 -i\nw
\end{pmatrix}
~.
%
\end{align}
\end{subequations}
The mode with $\lambda=-i\nw/2$ represents the incoming mode, and we choose it below.
There is no ambiguity for $\vecx_0$.
Once one obtains $\vecx_0$, $\vecx_n$ is obtained recursively:
\begin{align}
(\lambda+n-M_{-1})\vecx_n = \sum_{k=0}^{n-1} M_{n-1-k}\vecx_k~,
\quad(n\geq1)
\label{eq:recursion}
\end{align}
with $\lambda=-i\nw/2$. 
Using the recursion relation, one can find the coefficient vector $\vecx_1$:
\begin{subequations}
\label{eq:}
\begin{align}
%
%
\vecx_1 &= \frac{1}{4(\nw+i)}
\begin{pmatrix} 
2i(1-\nq^2)+\nw(1+2i\nw) \\ 
-4i-\nw(6+i\nw+2\nw^2)+2\nq^2(\nw+2i)
 \end{pmatrix}
\\
&=  -\frac{i(2\nq^2+1)}{4(\nw+i)}
\begin{pmatrix} 
1 \\ 
-1
 \end{pmatrix}
+(\text{regular})
 ~.
%
\end{align}
\end{subequations}
Here, we denote  regular expressions in $\nw$ as ``(regular)." 
Both components of $\vecx_1$ have the same poles. 
$\vecx_1$ becomes ambiguous at
\begin{align}
(\nw,\nq^2) = (-i, -1/2)~.
\label{eq:first_scalar}
\end{align}
This is a pole-skipping point. $\vecx_2$  is given by
\begin{align}
%
%
\vecx_2&=
 \begin{pmatrix}
  x_2^{(1)}\\
  x_2^{(2)}
\end{pmatrix}
=
\frac{i(2\nq^2+1)}{32(\nw+i)}
 \begin{pmatrix}
  2\nq^2-1\\
  -6\nq^2+7
\end{pmatrix}
+
\frac{i(\nq^2+1)(\nq^2+3)}{8(\nw+2i)}
 \begin{pmatrix}
  -1\\
  2
\end{pmatrix}
+(\text{regular})
~.
%
\end{align}
Both components of $\vecx_2$ become ambiguous at
\begin{align}
(\nw,\nq^2) =& (-i, -1/2)~, (-2i,-1)~, (-2i,-3)~.
%
\end{align}
$\vecx_2$ has pole-skipping points at $\nw=-i,-2i$.  The $\nw=-i$ pole-skipping point is the same as \eq{first_scalar}. 
Note that \textit{all residues of the coefficient vector should vanish at a pole-skipping point.} For example, the residue of $x_2^{(1)}$ vanishes at $(\nw,\nq^2)=(-i,1/2)$. But this is not a pole-skipping point because the residue of  $x_2^{(2)}$ does not vanish there. 
In such a case, the coefficient vector actually diverges.


\subsection{Maxwell scalar mode}\label{sec:maxwell_scalar}
\subsubsection{Maxwell scalar: master variable}\label{sec:maxwell_master}

As another example, consider the Maxwell field $A_M$. We consider the linear perturbation of the form $a_M e^{-i\omega t+iqx}$. The Maxwell scalar mode (diffusive mode) consists of $a_t,a_x,a_u$, but we use the $U(1)$ gauge-invariant variables:
\begin{subequations}
\label{eq:gauge_inv_variables}
\begin{align}
\mfa_t &= a_t+\frac{\omega}{q}a_x~, \\
\mfa_u &= a_u-\frac{1}{iq}a_x'~.
%
\end{align}
\end{subequations}
These gauge-invariant variables are nothing but field strength components: 
$F_{xt}=iq\mfa_t, F_{xu}=iq \mfa_u$.
The Maxwell equation $\nabla_N F^{MN}=0$ becomes
\begin{subequations}
\label{eq:maxwell}
\begin{align}
0 &= \mfa_t''+i\omega \mfa_u'-\frac{q^2}{4uf}\mfa_t~,
\label{eq:at} \\
0 &= (f \mfa_u)'+\frac{i\omega}{4uf}\mfa_t~, 
\label{eq:ax}\\
0 &= \mfa_u-\frac{i\omega}{\omega^2-q^2 f}\mfa_t'~.
\label{eq:au}
\end{align}
\end{subequations}
These 3 equations are not independent: 
 \eq{at} can be derived from the other 2 equations. 
 
The asymptotic behaviors of incoming modes are given by
\begin{subequations}
\label{eq:asymptotic}
\begin{align}
\mfa_t &\sim (1-u)^{-i\nw/2}~, \quad (u\to1)~. \\
\mfa_u&\sim (1-u)^{-1-i\nw/2}~, \quad (u\to1)~.
%
\end{align}
\end{subequations}
Note that $\mfa_t$ and $\mfa_u$ have different asymptotic behaviors.

Solve \eq{au} in terms of $\mfa_u$ and substitute it into \eq{ax}. One gets the master equation for $\mfa_t$:
\begin{align}
0&= f\mfa_t''+\frac{\nw^2f'}{\nw^2-\nq^2f} \mfa_t' +\frac{\nw^2-\nq^2f}{uf} \mfa_t~.
\label{eq:master1}
\end{align}

In this case, the system reduces to a single variable $\mfa_t$, so it is straightforward to apply the matrix formalism:
\begin{align}
\vecX &
= \begin{pmatrix} 
\mfa_t \\ 
f\mfa_t'
 \end{pmatrix}~.
%
\end{align}
The matrix $M$ is given by
\begin{align}
M &= 
  \begin{pmatrix} 
  0 & \dfrac{1}{f} \\
  \dfrac{ \stackrel{~}{-\nw^2+\nq^2f} }{uf} & \dfrac{\nq^2 f'}{-\nw^2+\nq^2 f}
  \end{pmatrix}~.
%
\end{align}
$M_{-1}$ is the same as \eq{M-1}, so its eigenvalues and eigenvectors are 
\begin{subequations}
\begin{align}
\lambda&=-i\nw/2~, \quad 
\vecx_0= 
\begin{pmatrix}
 1 \\
 i\nw 
\end{pmatrix}
~,
\\
\lambda&=+i\nw/2~, \quad 
\vecx_0= 
\begin{pmatrix}
 1 \\
 -i\nw
\end{pmatrix}
~.
%
\end{align}
\end{subequations}
%
%
There is no ambiguity for $\vecx_0$. 
$\vecx_1$ is given by
\begin{align}
%
%
\vecx_1&=
\frac{i\nq^2}{\nw}
 \begin{pmatrix}
  -1\\
  2
\end{pmatrix}
+
\frac{i(2\nq^2-1)}{4(\nw+i)}
 \begin{pmatrix}
  1\\
  -1
\end{pmatrix}
+(\text{regular})
~.
%
\end{align}
$\vecx_1$ becomes ambiguous at
\begin{align}
(\nw,\nq^2) =(0,0)~, (-i, 1/2)~.
\label{eq:diffusive}
\end{align}
Unlike the scalar field example, there is an additional ``hydrodynamic" pole-skipping point $(\nw,\nq^2) =(0,0)$ as well as the $\nw=-i$ pole-skipping point. This is because the field equation \eqref{eq:maxwell} or $M$ itself has the $0/0$ structure as $\nw,\nq\to 0$. 
In fact, $M_0$ is given by
\begin{align}
  M_0 &= 
   \begin{pmatrix} 
  0 & 0\\
0&  2\nq^2/\nw^2
  \end{pmatrix}
+(\text{regular})~,
\label{eq:M_0_maxwell}
\end{align}
so the right-hand side of the recursion relation \eqref{eq:recursion} is the origin of the hydrodynamic pole-skipping.
For the gravitational sound mode, the ``chaotic" pole-skipping at $\nw=+i$ also arises from the same reason. In any case, our formalism gives pole-skipping points both in the lower-half $\omega$-plane and in the upper-half $\omega$-plane unlike the conventional formalism \cite{Blake:2019otz}. 
The conventional formalism needs separate treatments for the ``chaotic" and ``hydrodynamic" pole-skippings.

Similarly, both components of $\vecx_2$ becomes ambiguous at
\begin{align}
(\nw,\nq^2) =& (0,0)~, (-i,1/2)~, (-2i,-1\pm\sqrt{3})~.
%
\end{align}

\subsubsection{Maxwell scalar: alternative master variable}\label{sec:alternative}

In previous subsection, we use $\mfa_t$ as the master variable. But one can choose $\mfa_u$ or a linear combination of $\mfa_t$ and $\mfa_u$ as the master variable. The choice of the master variable is not unique. This is sometimes problematic for a pole-skipping analysis because one cannot find all pole-skipping points if one does not choose an appropriate master variable. We illustrate this point using $\mfa_u$ as the master variable.%
\footnote{We pointed out this problem in the context of the AdS soliton background \cite{Natsuume:2023lzy}, but this is an important point, so we repeat the argument below. }
In this case, the master equation is given by
\begin{align}
0&= Z_u''+\left( \frac{f'}{f}+\frac{1}{u} \right) Z_u'+ \frac{\nw^2-\nq^2f}{uf^2} Z_u~,
\label{eq:master2}
\end{align}
where $Z_u=f\mfa_u$.
Choose $\vecX$ as 
\begin{align}
\vecX &
= \begin{pmatrix} 
Z_u \\ 
fZ_u'
 \end{pmatrix}~.
%
\end{align}
The matrix $M$ is given by
\begin{align}
M &= 
  \begin{pmatrix} 
  0 &\dfrac{1}{f} \\
  \dfrac{-\nw^2+\nq^2f}{uf} & -\dfrac{1}{u}
  \end{pmatrix}~.
%
\end{align}
$M_{-1}$ is the same as \eq{M-1}, so its eigenvalue and eigenvector for the incoming mode are 
\begin{align}
\lambda= -i\nw/2~,\quad
\vecx_0 =
  \begin{pmatrix}
 1\\
 i\nw
\end{pmatrix} 
~.
%
\end{align}
$\vecx_1$ is given by
\begin{align}
%
\vecx_1&=
\frac{i(2\nq^2-1)}{4(\nw+i)}
 \begin{pmatrix}
  -1\\
  1
\end{pmatrix}
+(\text{regular})
~.
%
\end{align}
Note that $\vecx_1$ has only the $\nw=-i$ pole and does not have the $\nw=0$ pole. 
$\vecx_1$ becomes ambiguous at
\begin{align}
(\nw,\nq^2) =(-i, 1/2)~.
%
\end{align}
Namely, the hydrodynamic pole-skipping point $(\nw,\nq^2) =(0,0)$ is missing in this variable. 
Put differently, the field equation \eqref{eq:master2} does not have the $0/0$ structure as $\nw,\nq\to0$.

Thus, the pole-skipping analysis based on a master variable has 2 problems:
\begin{enumerate}
\item
It is in general very difficult to find a master variable. Even its existence is not guaranteed. 
\item
The choice of a master variable is not unique. If one does not choose an appropriate master variable, one cannot find all pole-skipping points. To avoid this problem, one has to take into account all variables.
\end{enumerate}
Our matrix formalism is free from these problems.

\subsubsection{Maxwell scalar without master variables}\label{sec:maxwell_scalar_matrix}

For the Maxwell scalar mode, the master equation is available. But as an example of 2-field-system, let us directly consider \eq{maxwell}. In this case, both fields $\mfa_t, \mfa_u$ obey the first-order differential equations. So, it is easy to apply our matrix formalism. These fields have different asymptotic behavior \eqref{eq:asymptotic}. To take into account this point, choose $\vecX$ as
\begin{align}
\vecX = 
\begin{pmatrix} 
\mfa_t \\ 
f\mfa_u
 \end{pmatrix}
 ~.
%
\end{align}
The matrix $M$ and $M_{-1}$ are given by
\begin{subequations}
\begin{align}
M &= 
  \begin{pmatrix} 
  0 &2i\left(\dfrac{\nq^2}{\nw}-\dfrac{\nw}{f} \right) \\
  \dfrac{-i\nw}{2uf} &0
  \end{pmatrix}~,\\
M_{-1}&=i\nw
  \begin{pmatrix} 
  0& 1 \\
  1/4 &0 
  \end{pmatrix}~.
%
\end{align}
\end{subequations}
The eigenvalue and the eigenvector of $M_{-1}$ for the incoming mode are 
\begin{align}
%
%
\lambda&=-i\nw/2~, \quad 
\vecx_0= 
\begin{pmatrix}
 1 \\
 -1/2
\end{pmatrix}
~.
%
\end{align}
There is no ambiguity for $\vecx_0$. 
$\vecx_1$ is given by
\begin{subequations}
\begin{align}
%
%
\vecx_1&= 
 \begin{pmatrix}
  x_1^{(1)}\\
  x_1^{(2)}
\end{pmatrix}
=
-\frac{i\nq^2}{\nw}
 \begin{pmatrix}
  1\\
  0
\end{pmatrix}
+
\frac{i(2\nq^2-1)}{4(\nw+i)}
 \begin{pmatrix}
  1 \\
  1/2
\end{pmatrix}
+(\text{regular})
~.
\end{align}
\end{subequations}
Then, $\vecx_1$ becomes ambiguous at
\begin{align}
(\nw,\nq^2) =(0,0)~, (-i, 1/2)~.
%
\end{align}
Note that $x_1^{(1)}$ has the $\nw=0,-i$ poles, but $x_1^{(2)}$ has only the $\nw=-i$ pole. This implies that the hydrodynamic pole-skipping comes from $\mfa_t$ and explains why the hydrodynamic pole-skipping is missing in the analysis based on $\mfa_u$.

Similarly, $\vecx_2$ becomes ambiguous at
\begin{align}
(\nw,\nq^2) =& (0,0)~,  (-i,1/2)~, (-2i,-1\pm\sqrt{3})~.
%
\end{align}
Namely, we obtain the same results as the master variable one in \sect{maxwell_master}.

\subsection{Summary of the matrix formalism}\label{sec:summary}

It is now clear how to extend the formalism, and we summarize our formalism here:
\begin{enumerate}

\item
Suppose that a system has $m$ field equations that obey second-order differential equations. Rewrite them as $(2m)$ first-order differential equations. 
\item
Write them in the matrix form $0 = \vecX'- M \vecX$ with $(2m)\times(2m)$ component matrix $M$. Choose $\vecX$ so that all components have the same asymptotic behavior at the horizon $u\to1$. One can use the standard Frobenius method if the matrix $M$ behaves as
\begin{align}
M= \frac{M_{-1}}{u-1}+M_0+M_1(u-1)+\cdots~.
%
\end{align}

\item
Solve $\vecX$ by the Frobenius method:
\begin{align}
\vecX = \sum_{n=0}\, \vecx_n\, (u-1)^{n+\lambda}~.
%
\end{align}
The indicial equation for $\lambda$ is the eigenvalue equation for $M_{-1}$. Obtain the eigenvalues and the eigenvectors. There are $(2m)$ eigenvectors that   correspond to $m$ incoming and $m$ outgoing modes. Choose $m$ incoming modes $\vecx_{0,\eigen}~(\eigen=1,\cdots,m)$. 
\item
Using the recursion relation \eqref{eq:recursion}, find coefficient vectors $\vecx_{1,\eigen}$. This gives pole-skipping points at $\nw=-i$ as well as the ``hydrodynamic" and ``chaotic" pole-skipping points if there are any.
\item
Similarly, find higher vectors $\vecx_{n,\eigen}$. They give new pole-skipping points at $\nw=-in$. (See \sect{formal} for a formal proof.)
\item 
If one has a single field or a master variable, it is straightforward to obtain pole-skipping points. However, in our case, there are $m$ incoming eigenvectors, so the generic incoming eigenvector is a linear combination of these eigenvectors:
\begin{align}
\vecy_0 =C_1 \vecx_{0,1}+C_2 \vecx_{0,2}+\cdots+C_m\vecx_{0,m}~.
%
\end{align}
At pole skipping points, all residues of $\vecy_n$ should vanish. 
Obtain $C_\eigen$ and $\nq$ so that all residues vanish. \textit{Because we do not have a single master variable, this is the price we have to pay.} This last step is new and is explained below using  \HSCs.

\end{enumerate}
In \sect{formal}, we justify the above procedure by a formal argument.

\section{Pole-skipping for \HSCs}\label{sec:HSC_pole_skip}
\subsection{Holographic \SCs}

As a nontrivial example of the matrix formalism, we consider  \HSCs\ in the SAdS$_5$ \bh background:
\begin{align}
S_\text{m} &= -\frac{1}{g^2} \int d^5x \sqrt{-g} \biggl\{ \frac{1}{4}F_{MN}^2 + |D_M\Psi|^2+m^2|\Psi|^2\biggr\}~.
\label{eq:minimal}
\end{align}
Here, $D_M:=\nabla_M-iA_M$. We consider the probe limit where the backreaction of matter fields onto the geometry is ignored.

The bulk matter equations are given by
\begin{subequations}
\begin{align}
0 &= D^2\Psi-m^2\Psi~, \\
0 &= \nabla_NF^{MN} - J^M~,\\
J_M &= -i\{ \Psi^* D_M\Psi -\Psi(D_M\Psi)^*\} 
= 2\Im(\Psi^* D_M\Psi)~.
%
\end{align}
\end{subequations}

At high temperature, the bulk matter equations admit a solution:
\begin{align}
A_t=\mu(1-u)~,
A_i=0, \Psi=0~,
%
\end{align}
where $\mu$ is the chemical potential.
But the $\Psi=0$ solution becomes unstable at the critical point and is replaced by a $\Psi\neq0$ solution. Then, the bulk field $\Psi$ is dual to the order parameter $\psi$. 

A \HSC\ has 2 dimensionful control parameters $T$ and $\mu$, so the system is parameterized by a dimensionless parameter $\mu/T$. One can fix $T$ and vary $\mu$, or one can fix $\mu$ and vary $T$. We work in the unit $\pi T=1$, so we fix $T$.
The location of the critical point $\mu_c$ and the solutions in the low-temperature phase are usually not available, 
and one needs numerical computations. But the pole-skipping analysis itself does not require the explicit form of the solutions.

\subsection{High temperature phase}\label{sec:high_T}

In the high-temperature phase, $\Psi=0$, and the perturbation $\delta\Psi$ decouples from Maxwell perturbations. The pole-skipping of the Maxwell part is the same as the pure Maxwell case in \sect{maxwell_scalar}. 
The $\delta\Psi$-equation is given by
\begin{align}
0&=u \left( \frac{f}{u} \delta\Psi' \right)' + \left[ \frac{(2\nw+A_t)^2}{4uf} -\frac{\nq^2}{u} - \frac{m^2}{4u^2} \right] \delta\Psi~,
\label{eq:scalar_high}
\end{align}
where $A_t =\mu(1-u)$. For simplicity, we consider $m^2=-4$.
The field equation reduces to the minimally-coupled one when $A_t=0$. 
Then, the analysis is as simple as the minimally-coupled one, and one obtains
\begin{subequations}
\label{eq:high_T}
\begin{align}
(\nw, \nq^2) = \left(-i, \frac{-1-i\mu}{2} \right)~.
%
\end{align}
Similarly, the complex conjugate field $\delta\Psi^*$ satisfies \eq{scalar_high} with the replacement $A_t\to-A_t$ and has the pole-skipping point at 
\begin{align}
(\nw, \nq^2) = \left(-i, \frac{-1+i\mu}{2} \right)~.
%
\end{align}
\end{subequations}
At the critical point, the dual order parameter $\psi$ becomes massless and has a hydrodynamic pole in the sense $\nw,\nq\to0$. But there is no hydrodynamic pole-skipping if one approaches from the high temperature phase. 
This can be shown explicitly (\sect{analytic}).
But one also needs to approach from the low-temperature phase, where $\delta\Psi$ couples with $\delta\Psi^*$ and Maxwell perturbations.

\subsection{Low temperature phase}\label{sec:low_T}

\subsubsection{Matrix formalism}
In the low-temperature phase, $\Psi\neq0$. The background field equations are given by
\begin{subequations}
\label{eq:background}
\begin{align}
0 &= \bmA_t'' - \frac{1}{2u^2f}|\bmPsi|^2A_t~, \\
0 &= \left( \frac{f}{u} \bmPsi' \right)' + \left[ \frac{\bmA_t^2}{4u^2f} - \frac{m^2}{4u^3} \right] \bmPsi~. 
%
\end{align}
\end{subequations}
Here, boldface letters indicate the background solution. One can set $\bmPsi$ to be real.
In general, it is not possible to obtain analytic solutions for the backgrounds except the $m^2=-4$ case \cite{Herzog:2010vz}.

We consider the perturbations from the background $\bmPsi, \bmA_t$. We decompose $\Psi$ as the amplitude and its phase: 
\begin{align}
\Psi=\rho e^{i\theta} \to \delta\Psi= \delta\rho+ i\bmrho \delta\theta~.
%
\end{align}
We also use gauge-invariant variables. In the high-temperature phase, one can use gauge-invariant variables for Maxwell fields $\mfa_t,\mfa_u$. In addition, the following 2 variables are gauge-invariant in the low-temperature phase:
\begin{align}
\delta\rho~, \Theta := \delta\theta-\frac{1}{iq}a_x~.
%
\end{align}
We write perturbative equations using those gauge-invariant variables:
\begin{subequations}
\label{eq:eom}
\begin{align}
0 &= \mfa_t''+2i\nw\mfa_u' -\frac{2\nq^2u+\bmrho^2}{2u^2f}\mfa_t-\frac{\bmA_t\bmrho}{u^2f} \delta\rho
-\frac{i\nw\bmrho}{u^2f} \thetarho~,
\label{eq:at_HSC} \\
0 &= \mfa_t'+ \left\{ 2i\nw - \frac{if(2\nq^2u+\bmrho^2)}{u\nw} \right\} \mfa_u +\frac{if}{u\nw}\{\bmrho\thetarho'-\bmrho'\thetarho\}~,\\
0 &= (f\mfa_u)' + \frac{i\nw}{2uf}\mfa_t-
 \frac{\bmrho}{2u^2}\thetarho~, \\
0 &= u \left(\frac{f}{u}\delta\rho' \right)' +\frac{4u\nw^2+u\bmA_t^2-(m^2+4\nq^2u)f}{4u^2f}\delta\rho +\frac{\bmA_t\bm\rho}{2uf}\mfa_t+
\frac{i\nw\bmA_t}{uf}\thetarho~,
\\
0&= u \left(\frac{f}{u}\thetarho' \right)'+ \left( \frac{\bmrho}{u}-2\bmrho'\right)f \mfa_u - \frac{i\nw\bmA_t}{uf}\delta\rho
+\frac{4u\nw^2+u\bmA_t^2-(m^2+4\nq^2u+2\bmrho^2)f}{4u^2f} \thetarho~.
%
\end{align}
\end{subequations}
%
%
Just like the pure Maxwell case, not all equations are not independent from the other equations, so we do not use \eq{at_HSC} in the following analysis.

We use the following relation below:
\begin{subequations}
\begin{align}
\bmA_t(1) &=0~, \\
\bmrho'(1) &=-\frac{1}{8}m^2\bmrho(1)~.
%
\end{align}
\end{subequations}
The latter equation is derived from the background equation of motion \eqref{eq:background} near the horizon $u=1$.

We would like to choose $\vecX$ so that all components have the same asymptotic behavior at $u\to1$. For the Maxwell field, it is natural to choose $\mfa_t, f\mfa_u$ from \sect{maxwell_scalar_matrix}. For the complex scalar field, $\delta\rho,\delta\theta$ are related to the original variables $\delta\Psi,\delta\Psi^*$ as
\begin{align}
\delta\rho =\frac{\delta\Psi+\delta\Psi^*}{2}~, \quad
\bmrho\delta\theta =\frac{\delta\Psi-\delta\Psi^*}{2i}~,
%
\end{align}
so it is natural to choose $\delta\rho,\bmrho\Theta$. Namely, we choose $\vecX$ as
\begin{align}
%
{}^t\vecX = 
\begin{pmatrix}
\mfa_t & f\mfa_u & \delta\rho & f\delta\rho' & \bm\rho\Theta & f \thetarho' 
\end{pmatrix}
~.
%
\end{align}
In fact, all components have the same asymptotic behavior as we see below.
Then, the matrix $M$ diverges no more rapidly than $1/(1-u)$ as is evident from the perturbative equations \eqref{eq:eom}.

$M_{-1}$ is given by 
\begin{align}
M_{-1} &= \begin{pmatrix} 
0 & i\nw & 0 & 0 & 0 &0 \\ 
i\nw/4 & 0 & 0 & 0 & 0 & 0 \\
0 & 0 & 0 & -1/2 & 0 & 0 \\
0 & 0 & \nw^2/2 & 0 & 0 & 0 \\
0 & 0 & 0 & 0 & 0 & -1/2 \\
0 & 0 & 0 & 0 & \nw^2/2& 0 \\
 \end{pmatrix}~.
%
\end{align}
$M_{-1}$ has 3 eigenvectors with $\lambda=-i\nw/2$ and 3 eigenvectors with $\lambda=+i\nw/2$ .
This implies that our choice of $\vecX$ is appropriate. 
The eigenvectors that correspond to the incoming mode $\lambda=-i\nw/2$ are given by
\begin{subequations}
\begin{align}
 {}^t\vecx_{0,A} &= 
\begin{pmatrix}
1 & -1/2 &0 & 0 & 0 &0 &0
\end{pmatrix}
~, \\
 {}^t\vecx_{0,\rho} &=
\begin{pmatrix}
0 & 0 &1 & i\nw & 0 &0 
\end{pmatrix}
~,  \\
{}^t\vecx_{0,\theta} &= 
\begin{pmatrix}
0 & 0 &0 & 0 & 0 &1 &i\nw
\end{pmatrix}
~.
%
\end{align}
\end{subequations}
Because we do not use a master variable, there are 3 eigenvectors, so the generic eigenvector is given by
\begin{align}
\vecy_0 = C_A \vecx_{0,A}+C_\rho\vecx_{0,\rho}+C_\theta \vecx_{0,\theta}~,
%
\end{align}
where $C_A,C_\rho, C_\theta$ are constants that we determine below. There are 3 constants, but one constant is undetermined because we solve a linear perturbation problem. One chooses these constants for each pole-skipping point.

Below we choose $m^2=-4$ for simplicity, but the following analysis can be done for arbitrary $m^2$. We also express various expressions using $\bmrho(1),\bmA_t'(1)$, and we abbreviate them to $\bmrho,\bmA_t'$. For the $m^2=-4$ case, an analytic solution is available, but the solution is a perturbative expression in the condensate $\epsilon$ as we see below, and it is simpler to keep using $\bmrho,\bmA_t'$.

Using the recursion relation \eqref{eq:recursion}, obtain $\vecx_{1,\eigen}$. We first give only the pole structure:
\begin{subequations}
\begin{align}
%
%
{}^t\vecx_{1,A} &\sim\frac{1}{\nw+i}
\begin{pmatrix}
\dfrac{*}{\nw}& * &* & * & * &0 &0
\end{pmatrix}
+\text{(regular)}
~, \\
{}^t\vecx_{1,\rho} &\sim\frac{1}{\nw+i}
\begin{pmatrix}
0& 0 &* & * & * &* &*
\end{pmatrix}
+\text{(regular)}
~, \\
{}^t\vecx_{1,\theta} &\sim\frac{1}{\nw+i}
\begin{pmatrix}
*& * &* & * & * &* &*
\end{pmatrix}
+\text{(regular)}
~.
%
\end{align}
\end{subequations}
Here, we denote nonvanishing expressions as ``$*$". The first 2 components of $\vecx_{1,\rho}$ always vanish for a generic $m^2$. On the other hand, the last 2 components of $\vecx_{1,A}$ vanish only when $m^2=-4$. 
Among these vectors, only $\vecx_{1,A}$ has a pole at $\nw=0$ and can have a hydrodynamic pole-skipping.

\subsubsection{The hydrodynamic pole-skipping}

Because only $\vecx_{1,A}$ has a pole at $\nw=0$, it is enough to consider $\vecx_{1,A}$. Near $\nw=0$, 
\begin{align}
{}^t\vecx_{1,A} \sim \frac{1}{2\nw} 
\begin{pmatrix}
-i(2\nq^2+\bmrho^2)& 0 &0 & 0 & 0 &0 &0
\end{pmatrix}
+\text{(regular)}
~,
%
\end{align}
so the vector becomes ambiguous at
\begin{align}
(\nw,\nq^2)=\left(0,-\half\bmrho^2 \right)~.
%
\end{align}
At the critical point $\bmrho=0$, this is a hydrodynamic pole-skipping. 
But this corresponds to the Maxwell scalar pole-skipping \eqref{eq:diffusive} at $\omega=0$ since it comes from the $\mfa_t$ component. The order parameter becomes massless at the critical point, but there is no new hydrodynamic pole-skipping associated with the complex scalar field or the massless order parameter. We show this both in the high-temperature phase and in the low-temperature phase. Namely, not all hydrodynamic poles are pole-skipping points.

\subsubsection{The $\nw=-i$ pole-skipping}

All 3 vectors have poles at $\nw=-i$:
\begin{subequations}
\begin{align}
{}^t\vecx_{1,A} 
&\sim \frac{1}{\nw+i} 
\begin{pmatrix}
a_A & \half a_A &  b_A &  -b_A &  c_A &  -c_A
\end{pmatrix}
+\text{(regular)}~, \\
a_A &= \frac{i}{4} (2\nq^2-1+\bmrho^2)~, \\
b_A &= -\frac{i}{8}\bmrho \bmA_t'~, \\
c_A &= 0~.
%
\end{align}
\end{subequations}
\begin{subequations}
\begin{align}
{}^t\vecx_{1,\rho} &\sim \frac{1}{\nw+i} 
\begin{pmatrix}
a_\rho & \half a_\rho &  b_\rho &  -b_\rho &  c_\rho &  -c_\rho
\end{pmatrix}
+\text{(regular)}~, \\
a_\rho &=0~, \\
b_\rho &= -\frac{i}{4}(2\nq^2+1)~, \\
c_\rho &= \frac{i}{4}\bmA_t'~.
%
%
\end{align}
\end{subequations}
\begin{subequations}
\begin{align}
{}^t\vecx_{1,\theta} &\sim \frac{1}{\nw+i} 
\begin{pmatrix}
a_\theta & \half a_\theta &  b_\theta &  -b_\theta &  c_\theta &  -c_\theta
\end{pmatrix}
 +\text{(regular)}~, \\
a_\theta &=i\bmrho~, \\
b_\theta &= -\frac{i}{4} \bmA_t'~, \\
c_\theta &= -\frac{i}{4}(2\nq^2+1+\bmrho^2)~.
%
\end{align}
\end{subequations}
These vectors have 6 components, but only 3 components are independent and these vectors all take the form
\begin{align}
{}^t\vecx_{1} 
&\sim \frac{1}{\nw+i} 
\begin{pmatrix}
a & \half a &  b &  -b &  c &  -c
\end{pmatrix}
~.
\label{eq:relations}
\end{align}
Similar relations hold for $\vecx_{n,\eigen}$ at $\nw=-ni~(n>1)$.
 This becomes important below.
%
%

Now, consider the generic eigenvector $\vecy_0$. Then, $\vecy_1$ is given by
\begin{align}
\vecy_1 = C_A \vecx_{1,A}+C_\rho\vecx_{1,\rho}+C_\theta \vecx_{1,\theta}~.
%
\end{align}
In order to obtain pole-skipping points, all residues of  $\vecy_1$ should vanish at $\nw=-i$:
\begin{itemize}
\item
The vector $\vecy_1$ have 6 components, but there are only 3 independent components because of relations \eqref{eq:relations}.
\item
One can choose 2 constants among $C_A,C_\rho,C_\theta$.
\item
Then, one can make all residues to vanish by choosing $\nq^2$ appropriately. This gives pole-skipping points at $\nw=-i.$
\end{itemize}
Namely, one solves the following equations in terms of $C_A,C_\rho,C_\theta,\nq^2$: 
\begin{subequations}
\begin{align}
0 &=C_A a_A+C_\rho a_\rho+C_\theta a_\theta~,\\
0 &=C_A b_A+C_\rho b_\rho+C_\theta b_\theta~,\\
0 &=C_A c_A+C_\rho c_\rho+C_\theta c_\theta~.
%
\end{align}
\end{subequations}
In a matrix form,
\begin{align}
0&=R
\begin{pmatrix}
C_A \\
C_\rho \\
C_\theta
\end{pmatrix}
~, 
\quad
R =
\begin{pmatrix}
a_A & a_\rho &a_\theta \\
b_A & b_\rho &b_\theta \\
c_A & c_\rho &c_\theta 
\end{pmatrix}
~.
%
\end{align}
For the solution to exist, the residue matrix $R$ should satisfy
\begin{align}
\text{det}(R)=0~.
%
\end{align}
From the above explicit expressions of $a,b,c$, the determinant reduces to an $O(\nq^6)$ expression.
This means that there are 3 solutions of $\nq^2$. In this way, one can find 3 pole-skipping points $\nq^2$. 

One can obtain the exact expressions for these pole-skipping points, but they are complicated expressions. Instead, we consider the case $\bmrho\ll1$, namely near the critical point, and we give pole-skipping points at $O(\bmrho^2)$.

\begin{enumerate}
\item
One pole-skipping point corresponds to the Maxwell scalar pole-skipping \eqref{eq:diffusive}:
\begin{subequations}
\label{eq:first}
\begin{align}
\nq_1^2 &= \half + \frac{-4+\bmA_t'^2}{2(4+\bmA_t'^2)}\bmrho^2+\cdots~, \\
\frac{C_\rho}{C_A} &= -\frac{ \bmA_t'}{ 4+\bmA_t'^2} \bmrho+\cdots~, \\
\frac{C_\theta}{C_A} &=- \frac{ \bmA_t'^2}{ 2(4+\bmA_t'^2)} \bmrho+\cdots~.
%
\end{align}
\item
The other 2 pole-skipping points correspond to the complex scalar pole-skipping and its conjugate ones \eqref{eq:high_T}:
\begin{align}
\nq_2^2 &= -\half +\half i \bmA_t' - \frac{3\bmA_t'+2i}{ 4(\bmA_t'+2i) }\bmrho^2+\cdots~, \\
\frac{C_A}{C_\rho} &=  \frac{ 4\bmrho }{ \bmA_t'+2i}+\cdots~, \\
\frac{C_\theta}{C_\rho} &= -i-\frac{ \bmA_t'-2i }{ 2\bmA_t'(\bmA_t'+2i)}\bmrho^2+\cdots~.
%
\end{align}
$\nq_3^2$ is given by the complex conjugate of $\nq_2^2$:
\begin{align}
\nq_3^2 &= -\half - \half i \bmA_t' - \frac{3\bmA_t'-2i}{ 4(\bmA_t'-2i) }\bmrho^2+\cdots~, \\
\frac{C_A}{C_\rho} &= \frac{ 4\bmrho }{ \bmA_t'-2i}+\cdots~, \\
\frac{C_\theta}{C_\rho} &= i-\frac{ \bmA_t'+2i }{ 2\bmA_t'(\bmA_t'-2i)}\bmrho^2+\cdots~.
%
\end{align}
\end{subequations}
\end{enumerate}

The pole-skipping points at $\nw=-2i$ can be obtained in a similar manner. 

\subsubsection{Pole-skipping points by boundary quantities}\label{sec:analytic}

As mentioned earlier, there exists an analytic solution for the \HSC\ with $m^2=-4$ \cite{Herzog:2010vz}. Then,
\begin{enumerate}
\item 
One can rewrite pole-skipping points by boundary quantities, the condensate $\epsilon$ and the chemical potential $\mu$. 
\item
In the high-temperature phase, one can obtain the Green's function in the hydrodynamic limit exactly. One can check that there is no new hydrodynamic pole-skipping associated with the massless order parameter.
\end{enumerate}

Recall that we fix $T$ and vary $\mu$, so $\mu$ is the control parameter. In this case, the critical point is $\mu_c=2$. Near the critical point, the complex scalar field remains small, and one can expand matter fields. Namely, one can construct the low-temperature background perturbatively in $\epsilon$ where $\epsilon \ll 1$ is the condensate:
%
%
\begin{subequations}
\label{eq:herzog}
\begin{align}
\bmrho &= -\epsilon \frac{u}{1+u} +O(\epsilon^3)~, \\
\bmA_t &= 2(1-u)+ \epsilon^2 \left\{ \frac{1-u}{24} - \frac{u(1-u)}{4(1+u)} \right\}+O(\epsilon^5)~.
%
\end{align}
\end{subequations}
This gives
\begin{align}
\bmrho(1)=-\half\epsilon+\cdots~, \quad \bmA_t'(1)=-2+\frac{\epsilon^2}{12}+\cdots~.
%
\end{align}
Then, the $\nw=0$ pole-skipping point is
\begin{align}
(\nw,\nq^2)=\left(0,-\frac{\epsilon^2}{8}+\cdots \right)~.
%
\end{align}
This corresponds to the Maxwell scalar pole-skipping point \eqref{eq:diffusive} with the $O(\epsilon^2)$ correction.

The $\nw=-i$ pole-skipping points \eqref{eq:first} are
\begin{align*}
\nq_1^2 &= \frac{1}{2}+O(\epsilon^4)~,\\
\nq_2^2 &= -\half-i -\frac{6+i}{48}\epsilon^2+\cdots~, \\
\nq_3^2 &= -\half+i -\frac{6-i}{48}\epsilon^2+\cdots~.
%
\end{align*}

Near the critical point, the chemical potential is written as
\begin{align}
\mu=\mu_c+\epsmu~,
%
\end{align}
where $\epsmu$ is the derivation from the critical point. 
From \eq{herzog},
\begin{align}
\mu = \bmA_t|_{u=0} =2+\frac{\epsilon^2}{24}+\cdots~.
%
\end{align}
Then, the condensate $\epsilon$ and $\epsmu$ are related by 
\begin{align}
\epsilon^2=24\epsmu+O(\epsmu^2)~.
%
\end{align}
Finally, one can rewrite $\nq_2,\nq_3$ as
\begin{subequations}
\begin{align}
\nq_1^2 &= \frac{1}{2}+O(\epsilon^4)~,\\
\nq_2^2 &= \frac{-1-i\mu}{2} -\frac{1}{8}\epsilon^2+\cdots~, \\
\nq_3^2 &=  \frac{-1+i\mu}{2} -\frac{1}{8}\epsilon^2+\cdots~.
%
\end{align}
\end{subequations}
We rewrite $\nw=-i$ pole-skipping points by $\mu$ and $\epsilon$, but they are not independent: $\epsilon=\epsilon(\mu)$. 
The point $\nq_1$ corresponds to the Maxwell scalar pole-skipping point \eqref{eq:diffusive} with no correction at  $O(\epsilon^2)$. The points $\nq_2,\nq_3$ correspond to the complex scalar pole-skippings with the $O(\epsilon^2)$ corrections. In fact, when $\epsilon=0$, $\nq_2,\nq_3$ agree with the high-temperature pole-skipping points \eqref{eq:high_T}.

\paragraph{Green's function in high-temperature phase:}
In the high-temperature phase, one can obtain the Green's function (or the response function) in the hydrodynamic limit easily \cite{Natsuume:2018yrg}. 
Set $\epsmu\to l ^2\epsmu, q \to l q, \omega \to l^2\omega$, and expand $\delta\Psi$ as a series in $l$:
\begin{align}
\delta\Psi = (1-u^2)^{-i\omega/4} (F_1+ l^3 F_3+\cdots)~.
%
\end{align}
We impose the incoming-wave boundary condition at the horizon.
The solution is given by
\begin{subequations}
\begin{align}
F_1 &=- \Cone\,\, \frac{u}{1+u} \sim -\Cone\,  u~,
\quad(u\to0)~, \\
F_3 &= \Cone\frac{q^2-2\epsmu-(3+i)\omega}{8}\frac{u\ln u}{1+u} + \Cone\frac{(1-i)\omega+2\epsmu}{4} \frac{u\ln(1+u)}{1+u}~,
%
\end{align}
\end{subequations}
so the asymptotic behavior with $l\to1$ is given by 
\begin{align}
\delta\Psi &\sim \frac{1}{8}\Cone\{q^2-2\epsmu-(1-3i)i\omega\} u\ln u- \Cone u+\cdots~, 
\quad(u\to0)~.
%
\end{align}
%
%
From \eq{Green}, the Green's function is then given by
\begin{align}
 G^R_{\psi\psi} &=-\frac{\del \delta\psi}{\del J} = -\frac{4}{q^2-2\epsmu-(1-3i)i\omega}~.
\label{eq:high-T_order}
\end{align}
At the critical point $\epsmu\to0$, there is a hydrodynamic pole at $(\omega,q)=(0,0)$, but it is not associated with pole-skipping because the residue does not vanish.

\section{Formal analysis}\label{sec:formal}


For \HSCs, the generic coefficient vector $\vecy_1$ has 6 components, but there are only 3 independent components $a,b,c$ because of relations \eqref{eq:relations}. Also, $\vecy_1$ has 3 constants $C_\eigen$, and one can choose 2 constants among them. Then, one can make all residues to vanish by choosing $q$ appropriately. This must be the case if there is a pole-skipping point. But it is not clear if this always holds for any system. 
Here, we give a formal argument why this is true in general. 

Recall the eigenvalue equation \eqref{eq:eigen} and the recursion relation \eqref{eq:recursion} for the generic eigenvector $\vecy_0$:
\begin{subequations}
\begin{align}
M_{-1}\vecy_0 &=\lambda\vecy_0~,
\label{eq:recursion_rel-0th}
\\
(\lambda+n-M_{-1})\vecy_n &= \sum_{k=0}^{n-1} M_{n-1-k}\vecy_k~,
\quad(n\geq1)~.
%
\end{align}
\end{subequations}
Suppose that the $(2m)\times(2m)$ matrix $M_{-1}$ has $m$ eigenvalues $\lambda_\tin=-i\nw/2$ and $m$ eigenvalues $\lambda_\tout=i\nw/2=-\lambda_\tin$. The corresponding eigenvectors are denoted as $\vecx_{0,\eigen}^{\,\tin}$ and $\vecx_{0,\eigen}^{\,\tout}~(\eigen=1,\cdots,m)$, respectively.
We assume that these eigenvectors are linearly independent.%
 \footnote{See the remark at the end of this section.}
Then, 
\begin{center}
The generic coefficient vector $\vecy_n$ has only $m$ independent components  \\
that have poles at $\nw=-in$.
\end{center}
There are 2 implications:
\begin{itemize}
\item
First, the pole-skipping points at $\nw=-in$ first appear from $\vecy_n$ $(n\geq1)$. 
\item
Second, the generic eigenvector $\vecy_0$ has $m$ constants $C_\eigen$. One can choose $(m-1)$ constants among $C_\eigen$ and 1 constant $\nq$. Then, there are enough degrees of freedom to make all residues of $\vecy_n$ to vanish in principle. 
\end{itemize}
From our assumptions, there exists a $(2m)\times(2m)$ matrix $P$ such that 
\begin{align}
P^{-1}M_{-1}P
=
\begin{pmatrix}
\lambda_\tin \tI & \tO \\
\tO & -\lambda_\tin \tI
\end{pmatrix}
=: D~,
%
\end{align}
where $\tI$ and $\tO$ are $m\times m$ identity matrix and zero matrix.  Then, the recursion relation can be written as
\begin{subequations}
\label{eq:recursion_rel-matrix}
\begin{align}
(\lambda+n-D)(P^{-1}\vecy_n) &= 
\begin{pmatrix}
\up_n \\
\down_n
\end{pmatrix}
~,
\quad(n\geq1)~,
\\
\begin{pmatrix}
\up_n \\
\down_n
\end{pmatrix}
&:=\sum_{k=0}^{n-1} (P^{-1}M_{n-1-k}P)(P^{-1}\vecy_k) 
~.
%
\end{align}
\end{subequations}
Here, $\up_n,\down_n$ are $m$-dimensional vectors. 
For now, we assume that new pole-skipping points do not appear from $\up_n,\down_n$ although there are important exception as discussed below. 
If we choose the incoming mode $\lambda=\lambda_\tin$, 
\begin{align}
\lambda+n-D =
\begin{pmatrix}
(\lambda+n-\lambda_\tin) \tI & \tO \\
\tO & (\lambda+n+\lambda_\tin) \tI
\end{pmatrix}
=
\begin{pmatrix}
n \tI & \tO \\
\tO & (2\lambda_\tin+n) \tI
\end{pmatrix}
%
\end{align}
so that the recursion relation becomes
\begin{align}
\begin{pmatrix}
n \tI & \tO \\
\tO & (2\lambda_\tin+n) \tI
\end{pmatrix}
(P^{-1}\vecy_n)
=
\begin{pmatrix}
\vec{u}_n \\
\vec{d}_n 
\end{pmatrix}
~.
%
\end{align}
Then,
\begin{subequations}
\begin{align}
P^{-1} \vecy_n 
&=
\begin{pmatrix}
\dfrac{\vec{u}_n}{n} \\
\dfrac{ \stackrel{~}{\vec{d}_n} }{2\lambda_\tin+n}
\end{pmatrix}
~, \\
\to 
\vecy_n &= \frac{P}{n-i\nw}
\begin{pmatrix}
\vec{0}\\
\vec{d}_n 
\end{pmatrix}
+(\text{regular at }\nw=-in)~.
%
\end{align}
\end{subequations}
Therefore, $\vecy_n$ has only $m$ independent components that have poles at $\nw=-in$ $(n\geq1)$.
The left-hand side of the recursion relation \eqref{eq:recursion_rel-matrix} gives pole-skipping points only in the lower-half $\omega$-plane 
by construction. 

On the other hand, the``hydrodynamic" pole-skipping $(\nw=0)$ and the ``chaotic" pole-skipping $(\nw=+i)$ come from the right-hand side of the recursion relation \eqref{eq:recursion_rel-matrix}: they arise when $\vec{u}_n$ or $\vec{d}_n$ has the $0/0$ structure. For example, go back to the Maxwell scalar mode problem in \sect{maxwell_scalar}. For $n=1$, the right-hand side of the recursion relation gives $P^{-1}M_0~ \vecy_0$, but $M_0$ has the $0/0$ structure from \eq{M_0_maxwell}.

The explicit form of $P$ is given by
\begin{align}
P
=
\begin{pmatrix}
\vecx_{0,1}^{\,\tin}
&
\cdots
&
\vecx_{0,m}^{\,\tin}
&
\vecx_{0,1}^{\,\tout}
&
\cdots
&
\vecx_{0,m}^{\,\tout}
\end{pmatrix}
~.
%
\end{align}
We assume that these eigenvectors are linearly independent, so $P^{-1}$ exists.
For example, for the Maxwell scalar mode in \sect{maxwell_master},
\begin{align}
P=
\begin{pmatrix}
1 & 1 \\
i\nw & -i\nw
\end{pmatrix}
~.
%
\end{align}

The eigenvectors may not be independent, but this would happen at specific values of $(\omega,q)$. For the Maxwell scalar mode, $\det P=-2i\nw$, so it vanishes at $\nw=0$. This is not a problem. The Green's function is not uniquely determined at a pole-skipping point. This comes from the ``slope dependence" $\delta\omega/\delta(q^2)$ of the Green's function near the pole-skipping point. Namely, what we are really interested in is the Green's function near the pole-skipping point, not the one directly at the pole-skipping point. 

\section{Discussion}\label{sec:discussion}

\begin{itemize}

\item
In this paper, we propose a formalism to study the pole-skipping without relying on a master variable, and we apply it to  \HSCs. In general, ``hydrodynamic modes" are regarded  as pole-skipping points. For example, the Maxwell scalar mode has a hydrodynamic pole-skipping point $(\nw,\nq)=(0,0)$ in the high-temperature phase. However,
\begin{itemize}
\item
At the critical point, the order parameter (and its complex conjugate) become massless, and new hydrodynamic modes appear, but there is no new hydrodynamic pole-skipping associated with the massless order parameters.
\item
Meanwhile, there remains a pole-skipping point at $\nw=0$ associated with the Maxwell scalar mode, but $\nq^2\propto\epsilon^2$. Thus, as one deviates away from the critical point, there is no ``genuine" hydrodynamic pole-skipping point $(\nw,\nq)=(0,0)$ in the low-temperature phase (for the scalar mode).
\end{itemize}

\item
In this paper, we obtain pole-skipping points in the traditional sense. Namely, we search pole-skipping points by tuning $\omega$ and $q$ with a given $\epsilon$. But if one allows to choose a specific value of $\epsilon$, there may be more pole-skipping points. In other word, we search pole-skipping points in the $(\omega,q)$-plane, not in the $(\omega,q,\epsilon)$-plane.

\item
We consider the \HSC\ 
in the SAdS$_5$ background for simplicity. But it is straightforward to extend our analysis to the following systems, and they all have the same qualitative behaviors:
\begin{itemize}
\item
\HSCs\ with arbitrary bulk scalar mass $m^2$ in the SAdS$_{p+2}$ background.

\item
``Nonminimal" \HSCs\ in the SAdS$_5$ background \cite{Herzog:2010vz}:%
\footnote{We assume $K(1)=1+A|\Psi(1)|^2 \neq 0$.}
\begin{subequations}
\begin{align}
S_\text{m} &= -\frac{1}{g^2} \int d^5x \sqrt{-g} \biggl\{ \frac{1}{4}F_{MN}^2 + K|D_M\Psi|^2+V \biggr\}~,\\
K &= 1+A|\Psi|^2~,\quad
V= m^2 |\Psi|^2+B |\Psi|^4~.
%
\end{align}
\end{subequations}
$A$ and $B$ are bulk parameters. Our system \eqref{eq:minimal} is the ``minimal" \HSC\ with $A=B=0$.
When $m^2=-4$, an analytic solution is available for this system, so one can write results by boundary quantities like \sect{analytic}.

\item
\HSCs\ in the background of the form 
\begin{subequations}
\begin{align}
ds_5^2 &= \frac{1}{u} (-fdt^2+dx^2+dy^2+dz^2)+\frac{du^2}{4u^2f}~, \\
f  &\sim -2\pi T (u-1)+\cdots~,\quad 
\bmA_t \sim \bmA_t'(1)(u-1)+\cdots~,\quad 
(u\sim 1)~.
%
\end{align}
\end{subequations}
Here, $T\neq0$. For the SAdS$_5$, $f=1-u^2$.
\end{itemize}

\item
We consider holographic superfluids, but most results equally apply to holographic superconductors. The difference between two systems lies in the difference of the boundary conditions for the bulk Maxwell field at the asymptotic infinity $u\to0$:
\begin{itemize}
\item
For holographic superfluids, one imposes the Dirichlet boundary condition. As a result, the boundary Maxwell field is nondynamical and is added as an external source. 
\item
For holographic superconductors, one imposes the Neumann or the ``mixed" boundary conditions (see,\example, Ref.~\cite{Natsuume:2022kic}). In this case, the boundary Maxwell field is dynamical and the Higgs mechanism occurs on the boundary.
\end{itemize}
In \sect{analytic}, we impose the Dirichlet boundary condition for the bulk Maxwell field on the boundary, so it applies to a holographic superfluid. But the pole-skipping analysis itself is based on the near-horizon analysis, so the other results apply to holographic superconductors as well. 

\item
We do not discuss gravitational perturbations, but one can apply our formalism to gravitational perturbations, especially to the gravitational sound mode where the ``chaotic" pole-skipping point appears. One way to analyze the system is as follows:
\begin{itemize}
\item
In this problem, there appear 4 gauge-invariant variables which we denote as $\mfh_{tt},\mfh_{tu},\mfh_{uu}, \mfh_L$ (see, \example, Refs.~\cite{Natsuume:2019sfp,Natsuume:2023lzy}).
\item
The linearized Einstein equation contains 2 constraint equations without $u$-derivatives. 
One can eliminate 2 variables using the constraint equations. The resulting equations are 2 first-order differential equations for 2 variables. 
\item
Then, the analysis is similar to the pure Maxwell example in \sect{maxwell_scalar_matrix}, and one can find the chaotic pole-skipping.
\end{itemize}

\end{itemize}
We hope that our formalism will be helpful to analyze pole-skipping for the other systems where a master variable is not available. 

\section*{Acknowledgments}


This research was supported in part by a Grant-in-Aid for Scientific Research (25K07291) from the Ministry of Education, Culture, Sports, Science and Technology, Japan. 


\appendix

\section{Matrix formalism (with incoming-wave ansatz)}\label{sec:method2}

In the text, we impose the incoming-wave boundary condition by choosing the eigenvalue $\lambda=-i\nw/2$ (Method~1). 
But in the conventional pole-skipping analysis, one uses the Eddington-Finkelstein coordinates or imposes the incoming-wave boundary condition ansatz.
Namely, for the scalar field, set
\begin{align}
\phi =f^{-i\nw/2} \Zin
%
\end{align}
and study the field equation for $\Zin$. In this case, one chooses $\lambda=0$ in the recursion relation \eqref{eq:recursion}. This is the method that we developed in Ref.~\cite{Natsuume:2023lzy}  (Method~2).

It does not matter whether one works on $\phi$ or $\Zin$. 
But the details of the matrix formalism are slightly different, so it is worthwhile to summarize Method~2. 
We consider only the scalar field example below for simplicity.
 
The field equation for $\Zin$ typically takes the form 
\begin{align}
0= \Zin''+P(u)\Zin'+Q(u)\Zin~.
\label{eq:master_like}
\end{align}
$P$ and $Q$ are expanded as
\begin{align}
P &= \sum_{n=-1} P_n(u-1)^n~, \quad
Q= \sum_{n=-1} Q_n(u-1)^n~.
%
\end{align}
The field equation has a regular singular point at $u=1$, but $Q$ typically starts from $Q_{-1}$. Also, $P_{-1}=1-i\nw$ typically. 

Again, write the field equation in a matrix form:
\begin{subequations}
\begin{align}
0 &= \vecX'- M \vecX~, \\
\vecX &= \begin{pmatrix} 
\Zin \\ 
\Zin'
 \end{pmatrix}~,
\\
M &= 
  \begin{pmatrix} 
  0 &1 \\
  -Q & -P
  \end{pmatrix}~.
%
\end{align}
\end{subequations}
Here, we choose $\Zin$ and $\Zin'$ as $\vecX$ because $\Zin$ is regular at the horizon for the incoming-wave.
$M$ is expanded as \eq{M_exp}.
The solution can be written as a power series:
\begin{align}
\vecX = \sum_{n=0}\, \vecx_n\, (u-1)^{n+\lambda}~.
%
\end{align}
Substituting this into the field equation, at the lowest order, one obtains
\begin{align}
0 =(\lambda - M_{-1})\vecx_0~.
%
\end{align}
This is the indicial equation for $\lambda$ and is the eigenvalue equation for $M_{-1}$. 
The eigenvalue and the eigenvector of $M_{-1}$ are 
\begin{subequations}
\begin{align}
  & \lambda = 0~,
\quad \vecx_0
  = \begin{pmatrix} 
  1 \\
  -\dfrac{Q_{-1}}{P_{-1}}
    \end{pmatrix}~,
%
\\
  & \lambda = i\nw -1~,
\quad \vecx_0
  = \begin{pmatrix} 
  0 \\ 
  1 
  \end{pmatrix}
  ~.
\end{align}
\end{subequations}
Because we impose the incoming-wave ansatz, the mode $\lambda=0$ is the incoming mode, so we choose $\lambda=0$. Note that 2 components of $\vecx_0$ do not have the same poles.

When $m^2=-4$, the explicit form of $\vecx_0$ is
\begin{align}
\vecx_0= 
  \begin{pmatrix} 
  1 \\
  \dfrac{ b_1 }{4(\nw+i)}
  \end{pmatrix}
+(\text{regular})
~, \quad
b_1 =-i(2\nq^2+1)~.
\label{eq:vec0}
\end{align}
Then, $\vecx_0$ becomes ambiguous at 
\begin{align}
(\nw,\nq^2) =(-i,-1/2)~.
%
\end{align}
Using the recursion relation \eqref{eq:recursion}, $\vecx_1$ takes the form 
\begin{subequations}
\begin{align}
%
%
\vecx_1&=
 \begin{pmatrix}
  x_1^{(1)}\\
  x_1^{(2)}
\end{pmatrix}
=
\frac{b_1}{16(\nw+i)}
 \begin{pmatrix}
  4\\
  3-2\nq^2
\end{pmatrix}
+
\frac{b_2}{4(\nw+2i)}
 \begin{pmatrix}
  0\\
  1
\end{pmatrix}
+(\text{regular})
~,
\\
b_2&=-i(\nq^2+1)(\nq^2+3)~.
%
\end{align}
\end{subequations}
$\vecx_1$ becomes ambiguous at 
\begin{align}
 (\nw,\nq^2) =& (-i, -1/2)~,  (-2i,-1)~, (-2i,-3)~.
%
\end{align}
This gives the same result in \sect{formalism}. In Method 2, $\vecx_1$ has pole-skipping points at $\nw=-i,-2i$. 

Both in Method 1 and 2, all residues of $\vecx_n$ should vanish at a pole-skipping point. In this case, the residue of $x_1^{(2)}$ vanishes at $(\nw,\nq^2)=(-i,3/2)$. But this is not a pole-skipping point because the residue of $x_1^{(1)}$ does not vanish there. 
In such a case, $\vecx_1$ actually diverges.




There are both advantages and disadvantages whether one imposes the incoming-wave ansatz or not:
\begin{itemize}
\item
In Method~1, one can just use the original field equations which are simpler. In Method~2, the field equations become complicated especially when one has a multiple number of fields. 
\item
In Method~1, $\vecx_0$ is regular, and the ambiguity first appears at $\vecx_1$.
In Method~2, the ambiguity already appears at $\vecx_0$.
\end{itemize}


\footnotesize

\begin{thebibliography}{}


\bibitem{Maldacena:1997re}
  J.~M.~Maldacena,
  ``The Large N limit of superconformal field theories and supergravity,''
  Int.\ J.\ Theor.\ Phys.\  {\bf 38} (1999) 1113
   [Adv.\ Theor.\ Math.\ Phys.\  {\bf 2} (1998) 231]
  [hep-th/9711200].

\bibitem{Witten:1998qj}
  E.~Witten,
  ``Anti-de Sitter space and holography,''
  Adv.\ Theor.\ Math.\ Phys.\  {\bf 2} (1998) 253
  [hep-th/9802150].

\bibitem{Witten:1998zw}
  E.~Witten,
  ``Anti-de Sitter space, thermal phase transition, and confinement in gauge theories,''
  Adv.\ Theor.\ Math.\ Phys.\  {\bf 2} (1998) 505
  [hep-th/9803131].
  
\bibitem{Gubser:1998bc}
  S.~S.~Gubser, I.~R.~Klebanov and A.~M.~Polyakov,
  ``Gauge theory correlators from noncritical string theory,''
  Phys.\ Lett.\ B {\bf 428} (1998) 105
  [hep-th/9802109].


\bibitem{CasalderreySolana:2011us}
  J.~Casalderrey-Solana, H.~Liu, D.~Mateos, K.~Rajagopal and U.~A.~Wiedemann,
  \textit{Gauge/String Duality, Hot QCD and Heavy Ion Collisions} (Cambridge Univ.\ Press, 2014)
  [arXiv:1101.0618 [hep-th]].

\bibitem{Natsuume:2014sfa}
  M.~Natsuume,
  \textit{AdS/CFT Duality User Guide},
  Lecture Notes in Physics Vol. 903 (Springer Japan, Tokyo, 2015) 
  [arXiv:1409.3575 [hep-th]].

\bibitem{Ammon:2015wua}
  M.~Ammon and J.~Erdmenger,
  \textit{Gauge/gravity duality : Foundations and applications}
  (Cambridge Univ.\ Press, 2015).

\bibitem{Zaanen:2015oix}
  J.~Zaanen, Y.~W.~Sun, Y.~Liu and K.~Schalm,
  \textit{Holographic Duality in Condensed Matter Physics}
  (Cambridge Univ.\ Press, 2015).

\bibitem{Hartnoll:2016apf}
  S.~A.~Hartnoll, A.~Lucas and S.~Sachdev,
  \textit{Holographic quantum matter}
  (The MIT Press, 2018) 
  [arXiv:1612.07324 [hep-th]].

\bibitem{Baggioli:2019rrs}
M.~Baggioli,
\textit{Applied Holography: A Practical Mini-Course},
SpringerBriefs in Physics (Springer, 2019)
[arXiv:1908.02667 [hep-th]].

\bibitem{Kovtun:2004de}
P.~Kovtun, D.~T.~Son and A.~O.~Starinets,
``Viscosity in strongly interacting quantum field theories from black hole physics,''
Phys. Rev. Lett. \textbf{94} (2005), 111601
[arXiv:hep-th/0405231 [hep-th]].



\bibitem{Shenker:2013pqa}
  S.~H.~Shenker and D.~Stanford,
  ``Black holes and the butterfly effect,''
  JHEP {\bf 1403} (2014) 067
  [arXiv:1306.0622 [hep-th]].

\bibitem{Roberts:2014isa}
  D.~A.~Roberts, D.~Stanford and L.~Susskind,
  ``Localized shocks,''
  JHEP {\bf 1503} (2015) 051
  [arXiv:1409.8180 [hep-th]].

\bibitem{Roberts:2014ifa}
D.~A.~Roberts and D.~Stanford,
``Two-dimensional conformal field theory and the butterfly effect,''
Phys. Rev. Lett. \textbf{115} (2015) no.13, 131603
[arXiv:1412.5123 [hep-th]].

\bibitem{Shenker:2014cwa}
  S.~H.~Shenker and D.~Stanford,
  ``Stringy effects in scrambling,''
  JHEP {\bf 1505} (2015) 132
  [arXiv:1412.6087 [hep-th]].

\bibitem{Maldacena:2015waa}
  J.~Maldacena, S.~H.~Shenker and D.~Stanford,
  ``A bound on chaos,''
  JHEP {\bf 1608} (2016) 106
  [arXiv:1503.01409 [hep-th]].
  

\bibitem{Grozdanov:2017ajz}
  S.~Grozdanov, K.~Schalm and V.~Scopelliti,
  ``Black hole scrambling from hydrodynamics,''
  Phys.\ Rev.\ Lett.\  {\bf 120} (2018) no.23,  231601
  [arXiv:1710.00921 [hep-th]].
  
\bibitem{Blake:2018leo}
  M.~Blake, R.~A.~Davison, S.~Grozdanov and H.~Liu,
  ``Many-body chaos and energy dynamics in holography,''
  JHEP {\bf 1810} (2018) 035
  [arXiv:1809.01169 [hep-th]].


\bibitem{Grozdanov:2019uhi}
S.~Grozdanov, P.~K.~Kovtun, A.~O.~Starinets and P.~Tadi\'c,
``The complex life of hydrodynamic modes,''
JHEP \textbf{11} (2019), 097
[arXiv:1904.12862 [hep-th]].

\bibitem{Blake:2019otz}
M.~Blake, R.~A.~Davison and D.~Vegh,
``Horizon constraints on holographic Green's functions,''
JHEP \textbf{01} (2020), 077
[arXiv:1904.12883 [hep-th]].

\bibitem{Natsuume:2019xcy}
  M.~Natsuume and T.~Okamura,
  ``Nonuniqueness of Green's functions at special points,''
  JHEP {\bf 1912} (2019) 139
  [arXiv:1905.12015 [hep-th]].


\bibitem{Natsuume:2019sfp}
  M.~Natsuume and T.~Okamura,
  ``Holographic chaos, pole-skipping, and regularity,''
  PTEP {\bf 2020} (2020) no.1, 013B07
  [arXiv:1905.12014 [hep-th]].
  
\bibitem{Natsuume:2019vcv}
M.~Natsuume and T.~Okamura,
``Pole-skipping with finite-coupling corrections,''
Phys. Rev. D \textbf{100} (2019) no.12, 126012
[arXiv:1909.09168 [hep-th]].

\bibitem{Wu:2019esr}
X.~Wu,
``Higher curvature corrections to pole-skipping,''
JHEP \textbf{12} (2019), 140
[arXiv:1909.10223 [hep-th]].

\bibitem{Balm:2019dxk}
F.~Balm, A.~Krikun, A.~Romero-Berm\'udez, K.~Schalm and J.~Zaanen,
``Isolated zeros destroy Fermi surface in holographic models with a lattice,''
JHEP \textbf{01} (2020), 151
[arXiv:1909.09394 [hep-th]].

\bibitem{Ceplak:2019ymw}
N.~Ceplak, K.~Ramdial and D.~Vegh,
``Fermionic pole-skipping in holography,''
JHEP \textbf{07} (2020), 203
[arXiv:1910.02975 [hep-th]].

\bibitem{Liu:2020yaf}
Y.~Liu and A.~Raju,
``Quantum Chaos in Topologically Massive Gravity,''
JHEP \textbf{12} (2020), 027
[arXiv:2005.08508 [hep-th]].

\bibitem{Ahn:2019rnq}
Y.~Ahn, V.~Jahnke, H.~S.~Jeong and K.~Y.~Kim,
``Scrambling in Hyperbolic Black Holes: shock waves and pole-skipping,''
JHEP \textbf{10} (2019), 257
[arXiv:1907.08030 [hep-th]].

\bibitem{Ahn:2020bks}
Y.~Ahn, V.~Jahnke, H.~S.~Jeong, K.~Y.~Kim, K.~S.~Lee and M.~Nishida,
``Pole-skipping of scalar and vector fields in hyperbolic space: conformal blocks and holography,''
JHEP \textbf{09} (2020), 111
[arXiv:2006.00974 [hep-th]].

\bibitem{Abbasi:2020ykq}
N.~Abbasi and S.~Tahery,
``Complexified quasinormal modes and the pole-skipping in a holographic system at finite chemical potential,''
JHEP \textbf{10} (2020), 076
[arXiv:2007.10024 [hep-th]].

\bibitem{Jansen:2020hfd}
A.~Jansen and C.~Pantelidou,
``Quasinormal modes in charged fluids at complex momentum,''
JHEP \textbf{10} (2020), 121
[arXiv:2007.14418 [hep-th]].

\bibitem{Ramirez:2020qer}
D.~M.~Ramirez,
``Chaos and pole skipping in CFT$_{2}$,''
JHEP \textbf{12} (2021), 006
[arXiv:2009.00500 [hep-th]].

\bibitem{Ahn:2020baf}
Y.~Ahn, V.~Jahnke, H.~S.~Jeong, K.~S.~Lee, M.~Nishida and K.~Y.~Kim,
``Classifying pole-skipping points,''
JHEP \textbf{03} (2021), 175
[arXiv:2010.16166 [hep-th]].

\bibitem{Natsuume:2020snz}
M.~Natsuume and T.~Okamura,
``Pole-skipping and zero temperature,''
Phys. Rev. D \textbf{103} (2021) no.6, 066017
[arXiv:2011.10093 [hep-th]].

\bibitem{Kim:2020url}
K.~Y.~Kim, K.~S.~Lee and M.~Nishida,
``Holographic scalar and vector exchange in OTOCs and pole-skipping phenomena,''
JHEP \textbf{04} (2021), 092
[erratum: JHEP \textbf{04} (2021), 229]
[arXiv:2011.13716 [hep-th]].

\bibitem{Sil:2020jhr}
K.~Sil,
``Pole skipping and chaos in anisotropic plasma: a holographic study,''
JHEP \textbf{03} (2021), 232
[arXiv:2012.07710 [hep-th]].

\bibitem{Abbasi:2020xli}
N.~Abbasi and M.~Kaminski,
``Constraints on quasinormal modes and bounds for critical points from pole-skipping,''
JHEP \textbf{03} (2021), 265
[arXiv:2012.15820 [hep-th]].

\bibitem{Ceplak:2021efc}
N.~Ceplak and D.~Vegh,
``Pole-skipping and Rarita-Schwinger fields,''
Phys. Rev. D \textbf{103} (2021) no.10, 106009
[arXiv:2101.01490 [hep-th]].

\bibitem{Jeong:2021zhz}
H.~S.~Jeong, K.~Y.~Kim and Y.~W.~Sun,
``Bound of diffusion constants from pole-skipping points: spontaneous symmetry breaking and magnetic field,''
JHEP \textbf{07} (2021), 105
[arXiv:2104.13084 [hep-th]].


\bibitem{Natsuume:2021fhn}
M.~Natsuume and T.~Okamura,
``Nonuniqueness of scattering amplitudes at special points,''
Phys. Rev. D \textbf{104} (2021) no.12, 126007
[arXiv:2108.07832 [quant-ph]].

\bibitem{Blake:2021hjj}
M.~Blake and R.~A.~Davison,
``Chaos and pole-skipping in rotating black holes,''
JHEP \textbf{01} (2022), 013
[arXiv:2111.11093 [hep-th]].

\bibitem{Kim:2021xdz}
K.~Y.~Kim, K.~S.~Lee and M.~Nishida,
``Construction of bulk solutions for towers of pole-skipping points,''
Phys. Rev. D \textbf{105} (2022) no.12, 126011
[arXiv:2112.11662 [hep-th]].

\bibitem{Wang:2022mcq}
D.~Wang and Z.~Y.~Wang,
``Pole Skipping in Holographic Theories with Bosonic Fields,''
Phys. Rev. Lett. \textbf{129} (2022) no.23, 231603
[arXiv:2208.01047 [hep-th]].

\bibitem{Amano:2022mlu}
M.~A.~G.~Amano, M.~Blake, C.~Cartwright, M.~Kaminski and A.~P.~Thompson,
``Chaos and pole-skipping in a simply spinning plasma,''
JHEP \textbf{02} (2023), 253
[arXiv:2211.00016 [hep-th]].

\bibitem{Yuan:2023tft}
H.~Yuan, X.~H.~Ge, K.~Y.~Kim, C.~W.~Ji and Y.~Ahn,
``Pole-skipping points in 2D gravity and SYK model,''
JHEP \textbf{08} (2023), 157
[arXiv:2303.04801 [hep-th]].

\bibitem{Grozdanov:2023txs}
S.~Grozdanov and M.~Vrbica,
``Pole-skipping of gravitational waves in the backgrounds of four-dimensional massive black holes,''
Eur. Phys. J. C \textbf{83} (2023) no.12, 1103
[arXiv:2303.15921 [hep-th]].



\bibitem{Natsuume:2023lzy}
M.~Natsuume and T.~Okamura,
``Pole skipping in a non-black-hole geometry,''
Phys. Rev. D \textbf{108} (2023) no.4, 046012
[arXiv:2306.03930 [hep-th]].

\bibitem{Natsuume:2023hsz}
M.~Natsuume and T.~Okamura,
``Pole skipping as missing states,''
Phys. Rev. D \textbf{108} (2023) no.10, 106006
[arXiv:2307.11178 [hep-th]].

\bibitem{Abbasi:2023myj}
N.~Abbasi and K.~Landsteiner,
``Pole-skipping as order parameter to probe a quantum critical point,''
JHEP \textbf{09} (2023), 169
[arXiv:2307.16716 [hep-th]].

\bibitem{Ning:2023ggs}
S.~Ning, D.~Wang and Z.~Y.~Wang,
``Pole skipping in holographic theories with gauge and fermionic fields,''
JHEP \textbf{12} (2023), 084
[arXiv:2308.08191 [hep-th]].


\bibitem{Cartwright:2024iwc}
C.~Cartwright, U.~G{\"u}rsoy, J.~F.~Pedraza and G.~Planella Planas,
``Perturbing a quantum black hole,''
JHEP \textbf{03} (2025), 039
[arXiv:2408.08010 [hep-th]].


\bibitem{Grozdanov:2025ner}
S.~Grozdanov and M.~Vrbica,
``Duality and four-dimensional black holes: Gravitational waves, algebraically special solutions, pole skipping, and the spectral duality relation in holographic thermal CFTs,''
Phys. Rev. D \textbf{112} (2025) no.6, 066019
[arXiv:2505.14229 [hep-th]].

\bibitem{Lu:2025jgk}
Z.~Lu, C.~Ran and S.~f.~Wu,
``Bulk Spacetime Encoding via Boundary Ambiguities,''
Phys. Rev. Lett. \textbf{136} (2026) no.6, 061603
[arXiv:2506.12890 [hep-th]].

\bibitem{Ahn:2025exp}
Y.~Ahn, S.~Grozdanov, H.~S.~Jeong and J.~F.~Pedraza,
``Cosmological pole-skipping, shock waves and quantum chaotic dynamics of de Sitter horizons,''
[arXiv:2508.15589 [hep-th]].

\bibitem{Grozdanov:2025ulc}
S.~Grozdanov and M.~Vrbica,
``Thermal field theory correlators in the large-$N$ limit and the spectral duality relation,''
[arXiv:2509.18074 [hep-th]].

\bibitem{Davison:2025xdj}
R.~A.~Davison and H.~Jiang,
``Pole skipping from universal hydrodynamics of (1+1)d QFTs,''
[arXiv:2512.11024 [hep-th]].


\bibitem{Gubser:2008px}
S.~S.~Gubser,
``Breaking an Abelian gauge symmetry near a black hole horizon,''
Phys. Rev. D \textbf{78} (2008), 065034
[arXiv:0801.2977 [hep-th]].

\bibitem{Hartnoll:2008vx}
S.~A.~Hartnoll, C.~P.~Herzog and G.~T.~Horowitz,
``Building a Holographic Superconductor,''
Phys. Rev. Lett. \textbf{101} (2008), 031601
[arXiv:0803.3295 [hep-th]].

\bibitem{Hartnoll:2008kx}
S.~A.~Hartnoll, C.~P.~Herzog and G.~T.~Horowitz,
``Holographic Superconductors,''
JHEP \textbf{12} (2008), 015
[arXiv:0810.1563 [hep-th]].

\bibitem{Herzog:2010vz}
  C.~P.~Herzog,
  ``An Analytic Holographic Superconductor,''
  Phys.\ Rev.\ D {\bf 81} (2010) 126009
  [arXiv:1003.3278 [hep-th]].
  
\bibitem{Natsuume:2018yrg}
M.~Natsuume and T.~Okamura,
``Holographic Lifshitz superconductors: Analytic solution,''
Phys. Rev. D \textbf{97} (2018) no.6, 066016
[arXiv:1801.03154 [hep-th]].

\bibitem{Natsuume:2022kic}
M.~Natsuume and T.~Okamura,
``Holographic Meissner effect,''
Phys. Rev. D \textbf{106} (2022) no.8, 086005
[arXiv:2207.07182 [hep-th]].


\end{thebibliography}
\end{document}